\documentstyle[12pt]{article}

\title{Technical Notes      \\ on Complexity of the Satisfiability Problem} 

\author{Marek A. Suchenek}

\date{\scriptsize The Wichita State University     \\
KS 67208 U.S.A.}

\begin{document}

\bibliographystyle{alpha}

\newtheorem{prop}{Property}[section]
\newtheorem{ex}[prop]{Example}
\newtheorem{df}[prop]{Definition}
\newtheorem{probl}[prop]{Problem}

\maketitle


\bigskip


\begin{abstract}

These  notes contain, among others, a proof that the average running time of
an  easy  solution to the satisfiability problem for propositional calculus
is, under some reasonable assumptions, linear (with constant 2) in the size
of  the  input.  Moreover,  some  suggestions  are  made about criteria for
tractability  of  complex  algorithms. In particular, it is argued that the
distribution  of probability on the {\em whole} input space of an algorithm
constitutes  an  non-negligible factor in estimating whether the algorithm is
tractable or not.

\end{abstract}

\section{Introduction}
\label{Introduction}
\hspace*{\parindent}

It  is  not  unusual to hear computer professionals or students questioning
the  practical value of asymptotic complexity measures. To be honest, there
is  a  lot  of  evidence  of  occasional  discrepancy  between  algorithms'
asymptotic  and  actual  behaviors,  for example in the area of sorting and
multiplication.  After  all,  it  seems  typical,  that  authors (like Aho,
Hopcroft,  and  Ullman  in \cite{algorithms:AHU}, end of paragraph 1.4) rather
discourage the reader from drawing too many conclusions from the fact, that
a  running  time  of  an  algorithm  is  or  is  not  in  a certain $O$- or
$\Omega$-class.

\bigskip

The  theory  of  algorithms'  complexity is logical and clear: methods used
there do not seem to involve an unintentional error, since the authority of
mathematics  has  given  it  its  consent  of  the  thing  {\em  quod  erat
demonstrandum}. So, if it is so good then why it is so bad? To investigate
this  paradox  let  us  try to take a closer look at motivations of dealing
with asymptotic rather than actual complexities.

\bigskip

There  are  two of them: essential machine independence of algorithms to be
evaluated,  and  a virtual lack of limits on their inputs' size. The second
implies   an  expectation  that  inputs  may  grow  boundlessly,  which  is
considered  to  be  the reason for the most serious obstacle for successful
termination  of  a  run.  And  this  is why the asymptotic complexity of an
algorithm  has been supposed to characterize its performance on some future
inputs which (probably) may be brought to processing.

\bigskip

Depending  on  how  the  program's  complexity is expressed in terms of its
input's  size,  the  existing  asymptotic complexity measures fall into two
categories:  the  worst-case  and  the average one. The first kind seems to
reflect  an  implicit presumption of malicious gnome, who selects (possibly
the  most troublesome) inputs to the evaluated program. The second does not
allow  averaging  over  inputs  of  different  sizes, which causes at least
calculational  problems.  Both  of  them,  apparently  fully  adequate  for
evaluation  of  a  several  program,  may  completely  fail if applied to a
comparison:  program  {\em  A}  may  have  overall  better performance than
program {\em B}, with except for a few isolated ``worst" cases, when {\em B}
is much quicker than {\em A}; program {\em C} may have substantially better
average  efficiency  than  program  {\em  D}, however only for sufficiently
large  inputs  which  may  be not likely in all practical cases. Experience
shows, that the above scenarios are by no means artificial.

\bigskip

One  may suspect that dealing with future unforeseeable events may need some
probability  theory,  and  it is indeed the point of view which we advocate
here.  Because  in  estimating  how  much  time  will  be  spent  on future
computations one should take into account which elements of the input space
are  more  likely, and which are less. Moreover, since how one measures the
input  size  does  not seem to have much influence either the on actual nor
the  expected  running time of any particular program, we do not see a good
reason  why  expressing  averages  exclusively as a function of the size of
input  should be recognized as a universally satisfactory practice. (On the
contrary,  we  have  found it rather inadequate in our trials of evaluating an
average  running  time of the programs considered in this paper). Therefore
we  propose a modified notions of average running time and corresponding to
it $O$-classes.

\bigskip

In our approach we postpone abstraction from constant factors to some later
phase  of  evaluation.  In particular $ O(\bullet ^{2})$ and $ O(100 \times
\bullet  ^{2})$  classes  are  not identical in this paper. If one does not
need  to  deal  with  complexity  on  such  a  concrete  level, introducing
appropriate  equivalence  relation  (e.g.  one can impose $ O(\bullet ^{2})
\equiv  O(100  \times  \bullet  ^{2})  $),  will  easily translate obtained
results into a language of {\em modulo constants} complexity classes.

\bigskip

In   the   sequel,   we  will  use  an  NP-complete  problem,  namely:  the
satisfiability  problem  of propositional calculus, as one of illustrations
for  our  proposal.  Before doing this, we will start from some theoretical
considerations.

\bigskip

We   refer   the   reader   to   any   handbook  on  measure  theory  
for details concerning measure and probability spaces.
An extensive study of algorithms' complexity, including definitions of $O$-
and  $  \Omega$-classes,  satisfiability  problem,  NP-completeness theory,
NP-hard problems, and Cook's theorem, may be found in \cite{pap:comb}. Some
striking  results  about  {\em  better  than  expected} behavior of certain
algorithms    related    to    NP-hard    problems    may   be   found   in
\cite{
	Wilf84backtrack:an}.  Shannon's counting argument in the context of
complexity of Boolean functions appears in \cite{weg:comp}.

\section{Average running time, $O$-hierarchy, and tractability of
algorithms}
\label{Average}
\hspace*{\parindent}

Further on, we will use the von Neuman's definition of numbers, i.e. $0$ is
the  empty  set,  and $ n + 1 = n \cup \{ n \}\; (= \{ 0, ..., n \} ) $. We
denote  the  set  of  all  numbers  by $\omega$, and the set of all of them
without 0 by $ \omega^{+} $. Moreover, we apply $\bullet$ symbol to avoid
$ \lambda$ - expressions. Namely, $ f(\bullet) $ means $ \lambda x . f(x) $,
or in other words, $f$. E.g. $\bullet ^{3}$ denotes the cubic function.

\bigskip

We  will  fix  our attention on an algorithm $P$, with countable domain $X$
(which  we  will  call  the  input space), running time $ T : X \rightarrow
\omega^{+}$, distribution of probability $ \mu : X \rightarrow \langle 0, 1
\rangle  $,  extended  to  a normed measure $ \mu : {\cal P}(X) \rightarrow
\langle  0,  1  \rangle  $  by  $  \mu (Y) = \sum_{x \in Y} \mu(x) $
(usually, it is assumed that $ \mu (X) = 1) $.

\bigskip

By the average  running  time  $ T^{\mu}_{avg} $ (we use superscript $ \mu$ to
remind   the  explicit  role  of  the  probability  distribution  here)  we
understand a function defined for each $ Y \subseteq X $ as follows:

\medskip    

{\bf (i)} \hfill  $ T^{ \mu}_{avg} (Y) = \frac{\sum_{x \in Y} T(x) \times
\mu(x)}{\mu (Y)}  $. \hspace*{\fill}     \\
\smallskip     \\
It  is easily seen that the above expression defines the expected value of
$  T(x)  $  under  condition  $ x \in Y $, i.e. with respect to conditional
probability  $ \mu (x)/ \mu (Y) $. So its value tells us, how much time, on an
average,  the  algorithm  $  P  $ will spend running on a random input $x$,
provided it is known that $ x \in Y $.

\bigskip

If one would like to relate a running time to the size of input, a function
$  f : X \rightarrow \omega $, interpreted as an input size measure, comes handy.  It  partitions  the  input  space  onto  at most countably many
non-empty  subspaces,  which  are  abstraction  classes with respect to the
equivalence  relation  $  \equiv_{f} $ defined by: $ x \equiv_{f} y $ iff $
f(x)  = f(y) $. We will use $ X^{f}_{n} $ as an abbreviation for $ \{ x \in
X  \mid  f(x)  = n \} $ for any $ n \in \omega $. Under such conventions, a
relative  average  running  time  $  T^{f,  \mu}_{avg}$  of $ P$ is usually
defined by

{\bf (ii)} \hfill \[ T^{f, \mu}_{avg} (n) = \left\{ \begin{array}{ll}
1 & \mbox{if $\mu(X^{f}_{n})=0$}     \\
\frac{\sum_{x \in X^{f}_{n}} T^{f}_{x}(n) \times \mu (x)}{\mu (X^{f}_{n})}
& \mbox{otherwise},
\end{array}
\right. \]
    \\
where $ T^{f}_{x} $ satisfies for all $ x \in X $ (or at least for those
with $ \mu (x) \neq 0 $):
$ T^{f}_{x} (f(x)) = T(x) $.
One can see that for each $ n \in \omega $, such that $ \mu(X^{f}_{n})
\neq 0 $,
$ \sum_{x \in X^{f}_{n}} T^{f}_{x}(n) \times \mu (x) = 
\sum_{x \in X^{f}_{n}} T(x) \times \mu (x) $,
hence     \\
\[ T^{f, \mu}_{avg} (n) = T^{\mu}_{avg} (X^{f}_{n}). \]

Unlike  in  the  classic  case,  where averaging of related running time is
allowed  only  over abstraction classes $ X^{f}_{n} $, we admit the general
case,  i.e.  we  assume  that  the  relative  average  running  time may be
relativized  to  one  partition  of $X$, but averaged over another (one may
think:  orthogonal)  one.  However, instead of disentangling the dependency
between  an  average  value of $ T(x) $ and average size of $x$, which does
not  seem  simple, we will define directly, what it means that size-related
average  running  time  of  $P$ is in $ O(F)$-class for some function $ F :
\omega \rightarrow  \omega^{+}  $.  So,  let  $ \alpha : X \rightarrow
\omega $  be such a partition with corresponding abstraction classes $
X^{  \alpha}_{n}  $.  We say that $ T^{f, \mu}_{avg, \alpha} \in O(F) $ iff
for each $n$, such that $ \mu (X^{\alpha}_{n}) \neq 0 $,

\medskip    

{\bf (iii)} \hfill
  $  \sum_{x \in X^{ \alpha}_{n}} \frac{T(x)}{F(f(x))} \times \mu(x)
\leq \mu(X^{ \alpha}_{n}) $. \hspace*{\fill}     \\
\smallskip     \\
This means that the expected value of the quotient $
\frac{T(x)}{F(f(x))} $ over the set $ X^{ \alpha}_{n} $ does not exceed
1.

One  may  check  that  in the  usual case, where $ \alpha $ and $f$ coincide, $
T^{f, \mu}_{avg,f} \in O(F) $ iff for each $n$, such that $ \mu (X^{f}_{n})
\neq  0,  \: T^{\mu}_{avg} (X^{f}_{n}) \leq F(n) $, that is to say, $ T^{f,
\mu}_{avg}   (n)   \leq   F(n)  $.  Thus  our  definition  makes  a  proper
generalization of $O$-hierarchy of relative average running times.

\bigskip

It  is  not necessary that we understand $f$ as a measure of input size. We
may  think  of  $f$  as  the running time of another program $Q$ with input
space  $X$.  In light of such an interpretation, $ T^{f, \mu}_{avg, \alpha}
\in  O(F)  $  means  that  $F$ is an average upper bound of proportionality
factor  between  the  running time of $P$ and the running time of $Q$, over
each  class  $  X^{  \alpha}_{n}  $. E.g. if $c$ is a constant then $ T^{f,
\mu}_{avg,  \alpha}  \in  O(c  \times  \bullet)  $ means that in each $ X^{
\alpha}_{n} $, $Q$ is on average at most $c$ times quicker than $P$. If one
insists  on referring to the ordinary input's length, it may be measured by
the time which the simple rewriting program will spent on it.

\bigskip

Let  us  remind the reader here that our intention is, at least at earlier
stages of evaluation, not to abstract from the constant factor neglected in
the  classic  definition of $O$-hierarchy. This is why the coefficient at $
F(f(x))  $  in  (iii)  is  1.  Moreover, instead of dealing with asymptotic
behavior, we purposely introduced measures for the expected behavior, which
involves {\em all} possible inputs, so (iii) holds for all $ n'$s, not only
for those greater than some $ n_{0} $.

\bigskip

Finally,  we  define  the  notion  of algorithm's tractability. We call $P$
tractable over $ Y \subseteq X $ iff

\medskip    

{\bf (iv)} \hfill
$ T^{ \mu}_{avg} (Y)  < \infty $,
\hspace*{\fill}     \\
\medskip     \\
which means, that the expected length of the running time of $P$ is finite,
provided inputs are restricted to $Y$.

\bigskip

It  follows  from  the  above definition, that a linear algorithm (i.e. one
with  linear  running  time) may be not tractable at the same time, when an
exponential  one  is  tractable, however,  for  different probability distributions.  To  see  this  possibility,  let $ X = \omega, $ $ T(n) = n$, and
$S(n)  = 2^{n} $. If $ \mu(n) $ is proportional to $ n^{-2} $, and $ \nu(n)
$  to  $  2^{-2n}  $ then $ T^{ \mu}_{avg} ( \omega) = c \times \sum_{i \in
\omega}  \frac{1}{i}  =  \infty $ and $ S^{ \nu}_{avg} ( \omega) = d \times
\sum_{i \in \omega} 2^{-i} = 2d < \infty $.

\bigskip

One  may  notice,  that  in  the  definition of tractability, no input size
measure   is   explicitly   present.  This  is  consistent  with  a  simple
observation  that  how  long it takes to complete a run does not depend on
how  one  measures the size of the corresponding input. It should be noted,
however,  that  this natural from mathematical point of view definition may
be  somewhat impractical in certain cases. Clearly, if $ T^{ \mu}_{avg} (X)
=  \infty $ then you may expect the worst. But if not? Two statements $ T^{
\mu}_{avg}  (X) < $ 45 sec., and $ T^{ \mu}_{avg} (X) < $ 30,000 yrs., both
implying  the tractability of the program in question, have quite different
informational  content.  Because  in  our approach we did not abstract from
constant  factors  while  measuring program's complexity, our method may be
applied as well for evaluating the tractability in a stronger sense, where,
say, $ T^{ \mu}_{avg} (X) < $ 100 hrs. is required. It is quite clear, that
asymptotic  complexity  measures  do  not support, in general, this kind of
estimations.

\bigskip

If  one  is  interested  in  measuring  how  the  actual  running  time  is
distributed around its mean, other concepts of probability theory,
for instance, variance, or standard deviation, may be helpful.
We will not discuss them in this paper. Let us remark, however, that since 
$T(x)$ is a non-negative random variable,
 the probability that for $x \in Y$,
$  T(x)  \geq  \alpha  $  (where  $\alpha$ is a positive constant) does not
exceed    $    \frac{1}{\alpha    }    \times   T^{\mu}_{avg}(Y)   $.   
So, the computations longer than,
say, $100 \times T^{\mu}_{avg}(Y) $ will occur in $Y$ with at most $ 1 \% $
frequency.

\bigskip

For the sake of completeness of the picture we draw, let us state some basic
properties relating the introduced notions to each other.

\medskip    

\begin{prop}
\rm
\label{Property 1}

If $ T^{ \mu}_{avg} (X) < \infty $ then for every countable partition
$\alpha $ of input space $X$ on subsets of positive measure,

\medskip    

{\bf (v)} \hfill
$ T^{ \mu}_{avg} (X) = sup_{n \in \omega} T^{ \mu}_{avg} 
(\cup_{i \leq n} X^{ \alpha}_{i} ) $. \hspace*{\fill}     \\
\end{prop}
{\bf Proof}. By the definition, $ T^{ \mu}_{avg} (X) = \sum_{i \in \omega}  
\sum_{x \in X^{ \alpha}_{i}} T(x) \times \mu (x) $=     \\
\medskip \\
= $ \lim_{n \rightarrow \infty} \sum_{i \leq n} \sum_{x \in 
 X^{ \alpha}_{i}} T(x) \times \mu (x) = \lim_{n \rightarrow
\infty} \sum_{x \in \cup_{i \leq n} X^{ \alpha}_{i}}  T(x) \times \mu
(x) $ =     \\
\medskip \\
= $ \lim_{n \rightarrow \infty} T^{ \mu}_{avg} ( \cup_{i \leq n} X^{
\alpha}_{i} ) = ($ since $T(x) \times \mu (x) \geq 0) $ 
$ sup_{n \in \omega} T^{ \mu}_{avg} ( \cup_{i \leq n} X^{ \alpha}_{i})
$. \hfill $ \Box$

\begin{prop}
\rm
\label{Property 2}

Let  $ \mu $ be a normed measure on input space $ X $, 
let  $  \alpha$  be  a countable partition of  $X$ on subsets of
positive  measure, let $ f : x \rightarrow \omega $ be a measure of the
size  of  input,  and let $ F : \omega \rightarrow \omega^{+} $. In such 
circumstances 

\medskip    

{\bf (vi)} \hfill
$ T^{f, \mu}_{avg, \alpha} \in O(F) $
\hspace*{\fill}     \\
\medskip     \\
iff for each distribution $ \nu$ of probability satisfying

\medskip    

{\bf (vii)} \hfill
$  \nu(x) = c_{H} \times \frac{ H ( \alpha (x))}{F (f(x))} \times 
\mu (x) $, \hspace*{\fill}     \\
\medskip     \\
where $ H : \omega \rightarrow \omega $, the following inequality holds:

\medskip    

{\bf (viii)}  \hspace*{\fill}
$ T^{ \nu}_{avg} (X) \leq (F \circ f)^{\nu}_{avg} (X) $.
\hspace*{\fill}     \\

\end{prop}

\medskip    

{\bf Proof}. Let $ H : \omega \rightarrow \omega $. We have:
  \\
\medskip
  \\
$ T^{\nu}_{avg} (X) \leq ( F \circ f)^{\nu}_{avg} (X) \equiv $
$ \sum_{x \in X} T(x) \times \nu(x) \leq \sum_{x \in X} F(f(x)) \times \nu(x)
\equiv $
  \\
\medskip
  \\
$ \equiv \sum_{x \in X} T(x) \times c_{H} \times \frac{H( \alpha (x))}{F(f(x))}
\times \mu (x) \leq \sum_{x \in X} F(f(x)) \times c_{H} \times 
\frac{H( \alpha (x))}{F(f(x))} \times \mu (x) \equiv $
  \\
\medskip
  \\
$ \equiv \sum_{n \in \omega} \sum_{x \in X^{ \alpha}_{n}} T(x) \times c_{H} \times
\frac{H( \alpha(x))}{F(f(x))} \times \mu (x) \leq \sum_{n \in X} c_{H}
 \times H( \alpha (x)) \times \mu (x) \equiv $

\medskip    

{\bf (ix)} \hspace*{\fill}
$ \equiv \sum_{n \in \omega} H(n) \times 
 \sum_{x \in X_{n}^{\alpha}} \frac{T(x)}{F(f(x))}
\times \mu (x) \leq \sum_{n \in \omega} H(n) \times \mu (X^{ \alpha}_{n}) $.
\hspace*{\fill}     \\

If (vi) is true then by (iii) and (ix), we get (viii).

\bigskip

For proof of the converse implication let 
us assume (viii) and take as H in (vii) the characteristic
function of the set $ \{ m \} $, where $ m \in \omega $. In this case (ix)
may be reduced to \\

\smallskip

$ \sum_{x \in X^{ \alpha}_{m}} \frac{T(x)}{F(f(x))} \times \mu(x) \leq \mu
(X^{ \alpha}_{m}) $,
which gives (vi). \hfill $\Box$

\medskip    

Let us note here that constant $ c_{H} $ in (vii) is unambiguously determined 
by $H$, since  $ \nu(X) = 1 $.
Moreover, if $ f = \alpha $ then $ F(f(x)) $ in (vii) may be omitted.

\begin{prop}
\rm
\label{Property 3}

Let  $  \alpha$  be  a countable partition of input space $X$, 
let $ f : x \rightarrow \omega $ be a measure of the
size  of  input,  let  $ \mu $ be a  measure normed on each $ X^{ \alpha}_{n} 
$  (i.e. $ \mu (X^{ \alpha}_{n}) = 1$ for all $ n \in \omega) $, and let $
F, H : \omega \rightarrow \omega $. If for each $ n \in \omega $

\medskip    

{\bf (x)} \hfill
$ T^{f, \mu}_{avg, \alpha} \in O(F) $
\hspace*{\fill}     \\
\medskip     \\
then for every distribution $ \nu$ of probability satisfying

\medskip    

{\bf (xi)} \hfill
$ \nu(x) \leq \frac{ H( \alpha(x))}{F(f(x))} \times \mu (x) $ \hspace*{\fill}  
   \\
\medskip     \\
the following implication holds :

\medskip    

{\bf (xii)}  \hspace*{\fill}
  $  \sum_{n  \in  \omega}  H(n) < \infty \supset T^{ \nu}_{avg} (X) < \infty $.
\hspace*{\fill}     \\
\medskip     \\
\end{prop}

{\bf Proof}. (x) means that for each $ n \in \omega$:

\medskip    

{\bf (xiii)} \hspace*{\fill}
 $ \sum_{x \in X_{n}^{\alpha}} \frac{T(x)}{F(f(x))}
\times \mu (x) \leq 1 $.
\hspace*{\fill}     \\
\medskip     \\      
Hence $ T^{ \nu}_{avg} (X) = ($by (i) and $ \nu (X) = 1) \sum_{x \in X} 
T(x) \times \nu(x) \leq $     \\
\medskip   \\
= $ \sum_{n \in \omega} (\sum_{x \in X^{\alpha}_{n}} T(x) \times 
\frac{ \mu (x) \times H ( \alpha(x))}{F(f(x))}) $= 
$ \sum_{n \in \omega} (H(n) \times \sum_{x \in X^{\alpha}_{n}} 
\frac{T(x)}{F(f(x))} \times \mu (x)) \leq $ (by xiii)    \\
\medskip \\
$ \leq \sum_{n \in \omega} H(n) $, that is to say,
$ T^{ \nu}_{avg} (x) \leq \sum_{n \in \omega} H(n) $,
which gives us (xii). \hfill $\Box$

\bigskip

The  above  properties are useful in estimating tractability of algorithms.
Property  \ref{Property 1} gives us a tool for direct calculations of $ T^{
\mu}_{avg} (X) $. Using it one may also investigate the rate of growth of $
T^{  \mu}_{avg}  $  in function of $ \cup_{i < n} X^{\alpha}_{i} $, 
which may be useful if $
T^{  \mu}_{avg}  (X)  $  is  infinite,  or  finite but prohibitively large.
Putting  $  \alpha = f $ one can use known facts about average running time
in  classic  sense  in  estimating  the  tractability.  However,  it may be
somewhat difficult to discover a useful formula describing 
$ T^{ \mu}_{avg} ( \cup_{i < n} X^{f}_{n}) $. 
Property \ref{Property 2} allows estimations of tractability
in all cases the behavior of $F$ of is known.
Property  \ref{Property  3}  (being  as a matter of fact a
generalization  of  Property  \ref{Property 1}) may prove suitable in cases
Property \ref{Property 1} is not. It allows local analysis (i.e. in $ X^{
\alpha}_{n} $ subspaces) which using this property may be
extended to the whole input space.

\section{Complexity of tabulating program}
\label{Tabulating}
\hspace*{\parindent}

\medskip    

As  the  first  example  of  application  of  the  introduced notions, let us
evaluate   the   complexity  of  a  program,  which  given  a  sentence  of
propositional  calculus  tabulates  the  Boolean  function  defined by that
sentence. The problem of such tabulation is NP-hard.

\medskip

Even  relatively  simple  algorithms  (as one
rewriting  input  to  output)  may  be  intractable  if the distribution of
probability  does  not  decrease fast enough with the growth of input size.
Therefore  to  have  a  tractable instance of the problem one has to impose
some   conditions   on   rate   of   fading  of  probability  distribution.
Surprisingly, a relatively modest condition will suffice for this end.

\medskip

We will start from input space $X$ containing binary representations (using
e.g.  ASCII  or EBCDIC codes) of all propositional sentences in the reverse
Polish  form,  which  are  composed  of  some  countably  infinite  set  of
propositional  variables, and any complete set of logical connectives (e.g.
$  \vee, \wedge,$ and $ \neg$). As input size measure $f$ we will adopt the
length  (in  bits) of the representation mentioned above. As the orthogonal
partition $ \alpha $ we will use the number $ \alpha (x) $ of propositional
variables  appearing  in  the input $ x \: ( \alpha $ and $f$ are not fully
independent,  since $ f(x) $ cannot be less than $ \alpha (x)$; we will not
use  this  fact,  however).  We  will  assume, that the running time of the
program  for  any  input  $x$  is  equal to $ 2^{ \alpha(x)} \times f(x) $,
measured  in  some  abstract units of time. It is quite obvious, that there
exists  an  algorithm returning this ``efficiency": if it runs too fast, it
delays in printing the answer until the time $ 2^{ \alpha(x)} \times f(x) $
will  have  been  exhausted. Of course, one can probably construct a faster
program,  but  this  one  will  suffice  for  our  purposes.  It is perhaps
paradoxical,  nevertheless  clear,  that  only  the tiny inputs are causing
problems  with relative efficiency of our algorithm, since for large inputs
$x$  of  size  greater  than  $ 2^{ \alpha (x)} $ it has quite good, linear
performance.  On  the  other  hand,  the  number  of  such  tiny  inputs is
relatively so small in comparison  to  the  number  of  all  non-equivalent
propositions  of minimal lengths that it may be unable to lead us away from
polynomial average hierarchy.

\bigskip

We  will split each $ X^{ \alpha}_{n} $ (the set of all propositions of $X$
with $n$ propositional variables) onto a family of its subsets $ Y^{n}_{0},
Y^{n}_{1},  ...,  Y^{n}_{i},  ...,  $ so that $ Y^{n}_{0} $ will consist of
some sort of shortest sentences of $ X^{ \alpha}_{n}, \: Y^{n}_{1} $ of the
same  sort  of sentences of $ X^{ \alpha}_{n} \setminus Y_{0}, $ and so on.
Namely,  we define a function $ min : {\cal P}(X) \rightarrow {\cal P}(X) $
by:

\medskip    

{\bf (i)} for every element of $ Y \subseteq X $ there exists a logically
equivalent to it element of $ min(Y) $

\medskip

{\bf (ii)} for every element $x$ of $ min(Y) $ and every element $ y$ of $Y$, if
$x$ is logically equivalent to $y$ then $ f(x) \leq f(y) $

\medskip

{\bf (iii)} $ min(Y) $ is a minimal set satisfying (i) and (ii).

\medskip

To  demonstrate  the existence of such $ min(Y)$ one has to make use of the
axiom of choice: from each class of abstraction for the logical equivalence
on  $Y$ pick up an element $x$ with possibly smallest value of $ f(x)$. The
set constructed this way happened to automatically satisfy condition (iii).

\bigskip

Now  for  each  $ n \in \omega$ pose $ Y^{n}_{0} = min (X^{ \alpha}_{n}) $,
and $ Y^{n}_{i+1} = min(X^{ \alpha}_{n} \setminus \cup_{j \leq i} Y^{n}_{j}
) $. Of course, we have

\medskip

{\bf (iv)} \hspace*{\fill}
 $ X^{ \alpha}_{n} = \cup_{i \in \omega} 
Y^{n}_{i} $, \hspace*{\fill}     \\
\medskip     \\
and  for any $ i \neq j, \: Y^{n}_{i} \cap Y^{n}_{j} = 0 $. Let us estimate
lower  bounds  for  lengths  of  codes  of elements in $ Y^{n}_{i}$. Each $
Y^{n}_{i}  $  contains  the  number of elements equal to the cardinality of
Lindenbaum's  algebra  with  $n$ generators, or - equivalently - of Boolean
algebra of functions with $n$ variables, that is to say, $ 2^{2^{n}} $. Let
us  assume,  that probability distribution $ \mu$ assigns the same value to
all  elements of $ Y^{n}_{i} $. A semantical argument of 1-1 correspondence
between  the  elements  of  $  Y^{n}_{i}  $ and elements of mentioned above
algebras  shows,  that  this assumption is reasonable. It will enable us to
apply Shannon's counting argument.

\bigskip

To  evaluate  the  value  of $ \sum_{x \in Y^{n}_{i}} \frac{T(x)}{f^{3}(x)}
\times  \mu(x)  $, equal to $ \sum_{x \in Y^{n}_{i}} \frac{2^{n}}{f^{2}(x)}
\times  \mu(x)$,  let  us  observe that for every function $ g : \omega
\rightarrow  \omega $  such that for all $ x, \: g(x) \leq f(x) $, the
inequality  $  \sum_{x  \in Y^{n}_{i}} \frac{2^{n}}{f^{2}(x)} \times \mu(x)
\leq  \sum_{x  \in Y^{n}_{i}} \frac{2^{n}}{g^{2}(x)} \times \mu(x) $ holds.
Therefore  we  may  safely  assume  that  each $ Y^{n}_{i} $ contains all $
2^{2^{n}}$  shortest  binary  codes, giving the absolute lower bound of $f$
for all $Y^{n}_{i}$ together. In this case $ Y^{n}_{i} $ is composed of all
the codes of length $ \leq 2^{n} - 1$, and of one code of length $ 2^{n} $.

\medskip

We have: 

\medskip

$ \sum_{x \in Y^{n}_{i}} \frac{T(x)}{f^{3}(x)} \times \mu(x) $ = 
$ \sum_{x \in Y^{n}_{i}} \frac{2^{n}}{f^{2}(x)} \times \mu(x) $ = 
$ 2^{n} \times \mu(y_{0}) \times \sum_{x \in Y^{n}_{i}} \frac{1}{f^{2}(x)} $ = 
    \\
\medskip 
   \\
= $ 2^{n} \times \mu(y_{0}) \times ( \sum^{2^{n}-1}_{i=1} 
\frac{1}{i^{2}} \times 2^{i} + \frac{1}{(2^{n})^{2}} ) \leq $
$  2^{n}  \times  \mu(y_{0}) \times \sum^{2^{n}}_{i=1} \frac{1}{i^{2}}
\times 2^{i} \leq 2^{n} \times \mu(y_{0}) \times 2^{n} \times
\frac{1}{(2^{n})^{2}} \times 2^{2^{n}} =$
    \\
\medskip
  \\
= $ \mu(y_{0}) \times 2^{2^{n}} = \mu(Y^{n}_{i}) $, 
where $y_{0}$ is any element of $ Y^{n}_{i} $.
From (iv) follows
    \\
\medskip
  \\
$ \sum_{x \in X^{ \alpha}_{n}} \frac{T(x)}{f^{3}(x)} \times
\mu(x) $ 
= $ \sum_{i \in \omega} \sum_{x \in Y^{n}_{i}} 
\frac{T(x)}{f^{3}(x)} \times \mu(x) $ 
 $ \leq \sum_{i \in \omega} \mu(Y^{n}_{i})
=  \mu(\cup_{i \in \omega}  Y^{n}_{i})
 = \mu(X^{ \alpha}_{n}) $.

\medskip

According  to our definition of $O$-class, it means that $ T^{f, \mu}_{avg,
\alpha}  \in  O(\bullet  ^{3})  $. 
Applying  Property  \ref{Property 3} and taking into account $ \alpha(x)
\leq f(x) $ we conclude  that  if  for every $ x$, 

$\nu(x) \leq c \times \frac{ \mu(x)}{f^{d}(x) \times \mu(X^{ \alpha}_{n})}$, 
where $ d > 4$, then $ T^{ \nu}_{avg} (X) < \infty $.

\bigskip

We  were  not able to draw this conclusion
using  exclusively  Property  \ref{Property  1},  which  suggests  that our
generalized notion of average running time $O$-hierarchy may be more useful
that the classic one.

\medskip    

\section{Complexity of the satisfiability problem}
\label{Satisfiability}
\hspace*{\parindent}

\medskip    

The satisfiability problem of propositional calculus may be formulated as
follows.

\begin{quote}

Given a sentence $ \varphi$ of propositional calculus, decide whether there
exists  a  truth-valued  assignment for its proportional variables making $
\varphi $ true.

\end{quote}

All  known  deterministic  solutions  to  the satisfiability problem are of
exponential  worst-case time complexity. However, the question of existence
of polynomial solution still remains open. If the answer is ``yes" then, as
Cook has shown, every problem, which may be non-deterministically solved in
polynomial  worst-case  time,  can  also  be  solved  deterministically  in
polynomial worst-case time. This is the celebrated P=NP problem.

\bigskip

Instead  of  investigating  the  worst-case  running  time  of the quickest
solution  of  the  satisfiability  problem,  we  will   answer  more
practical question of its tractability, instead. A positive result we have been able
to  achieve in this respect makes, in our opinion, the P=NP problem slightly
less dramatical.

\bigskip

One  may  expect,  that  testing the satisfiability should be easier than
tabulating  a Boolean function. Indeed, for all but unsatisfiable sentences
(describing  the  constant  false Boolean function) one may stop trying all
possible assignments after the first satisfying one has been found. Now our
program  will  stop  either  if  it  found  an  assignment making its input
sentence true or if it examined unsuccessfully all possible assignments.

\bigskip

How  much  time  will it save us on average? We will show that surprisingly
much, as it follows from an elementary property of subsets of the set $ M
= \{0, ..., M-1\} $: assuming fair distribution of probability on ${\cal P}
(M)$,  the  expected  value of minimal element in a random subset of $M$ 
(which is the same as the expected number of tosses of a coin until {\em
heads} appears) is
less  than  2, no matter how large is $M$. Qualitatively similar observation one
can find in \cite{Wilf84backtrack:an}, pages 216--221, where the author proves that
the  average  number  $N$  of  nodes in the backtrack search tree of a {\em
random}  graph  subjected  to  coloring  with  at  most  $n$  colors may be
approximated  regardless  of the size of the graph; e.g. if $n = 3$ then $N
\approx 197$.

\bigskip

With each proposition $ \varphi_{n} $ of $n$ proportional variables we will
associate  its  model:  a set ${\cal K}( \varphi_{n}) $ of all assignments,
coded  as  binary sequences of length $n$, which make $ \varphi_{n} $ true.
Since  every such sequence constitutes a number from the interval $ \langle
0, 2^{n}-1 \rangle $, models may be thus understood as subsets of $ 2^{n} =
\{  0, ..., 2^{n}-1 \} $. We assume that the program testing satisfiability
scans  all  numbers  $m$  from  0  to  $ 2^{n-1} $, verifying for each $m$,
whether  its binary representation satisfies a sentence in question or not.

\bigskip

The  time  (measured  in some abstract units) our program will spent on any
input $x$ with $n$ propositional variables is given by:

\[ T(x) = f(x) \times (min_{n} ({\cal K}(x)) + 1) \]

where 

\[ min_{n} ({\cal K}) = \left\{ \begin{array}{ll}
2^{n} & \mbox{iff ${\cal K}=0$}     \\
min({\cal K}) & \mbox{otherwise}
\end{array}
\right. \]

\smallskip

Having  a  model ${\cal K} $ one may think of the set of all propositions $
\varphi_{n}  $,  for  which ${\cal K} $ is the model. Let us denote it by $
Th({\cal   K})   $.   Using  similar  semantical  argument  as  in  section
\ref{Tabulating},  we  assume  that  given $n$, it is equally likely that a
random formula $ \varphi $ falls in any class $ Th({\cal K}) $. In terms of
probability distribution $ \mu $ it means that for each $n$ and every two $
{\cal  K}, {\cal L} \subseteq 2^{n}, \mu(X^{ \alpha}_{n} \cap Th({\cal K}))
= \mu(X^{ \alpha}_{n} \cap Th({\cal L})). $

\bigskip

We have:
    \\
\medskip
    \\
$ \sum_{x \in X^{ \alpha}_{n}} \frac{T(x)}{2 \times f(x)} \times \mu(x) $ 
= $ \sum_{{\cal K} \subseteq 2^{n}} 
\sum_{x \in X^{ \alpha}_{n} \cap Th({\cal K})} \frac{T(x)}{2 \times f(x)} 
\times \mu(x) $ =     \\
\medskip     \\
= $ \sum_{{\cal K} \subseteq 2^{n}}  
\sum_{x \in X^{ \alpha}_{n} \cap Th({\cal K})} 
\frac{min_{n}({\cal K}) + 1}{2} \times \mu(x) $ 
= $ \sum_{{\cal K} \subseteq 2^{n}}  
 \frac{min_{n}({\cal K}) + 1}{2} 
\sum_{x \in X^{ \alpha}_{n} \cap Th({\cal K})} \mu(x) $ =     \\
\medskip
    \\
= $ \sum_{{\cal K} \subseteq 2^{n}} \frac{min({\cal K}) + 1}{2}
\times \mu (X^{ \alpha}_{n} \cap Th({\cal K})) $ 
= $ \sum_{{\cal K} \subseteq 2^{n}}  
\frac{min({\cal K}) + 1}{2} \times \frac{ \mu(X^{ \alpha}_{n})}
{2^{2^{n}}} $ = 
    \\
\medskip
    \\
= $ \frac{ \mu(X^{ \alpha}_{n})}{2} \times \sum_{{\cal K} \subseteq 2^{n}} 
\frac{min_{n}({\cal K}) + 1}{2^{2^{n}}} $ 
= $ \frac{ \mu(X^{ \alpha}_{n})}{2} \times \sum^{2^{n}}_{i=1}  
\frac{i \times 2^{2^{n}-i}}{2^{2^{n}}} $
$ < \frac{ \mu(X^{ \alpha}_{n})}{2} \times \sum^{ \infty}_{i=1}  
\frac{i}{2^{i}} = \mu(X^{ \alpha}_{n})$. 
\\
\medskip
\\
Hence $ T^{f, \mu}_{avg, \alpha} \in O(2 \times \bullet) $.

\bigskip    

Applying Property \ref{Property 3} we conclude that if for every 
$x$, 

$ \nu (x) \leq c \times \frac{\mu(x)}{f^{d}(x) \times \mu(X^{\alpha}_{n})}$,
where $ d>2$, then $ T^{\nu}_{avg}(X) < \infty  $. Again we were not lucky
enough to get the same
result using classic complexity measures.

\bigskip

The  same  calculations  prove  the  above  for  the  co-problem.  Also, the
NP-completeness of the satisfiability problem seems to be a rich source of
similar  estimations  for  other known complex problems. E.g. the mentioned
above  graph  coloring  with  backtrack  search,  or simplex algorithm (see
\cite{Wilf84backtrack:an}  for  its  analysis)  have been known to have better than
exponential   average  performance.  

\section{Higher order moments}
\label{Higher}

\hspace*{\parindent}

Similar calculations show that the $m$-th moment of $T(x)$, that is to say,
the average $m-$th power of the running time of the program mentioned in 
section \ref{Satisfiability}
 is in 
\\
$O(2.5 \times m^{m+1} \times \bullet ^{m})$. Namely,
for $m \geq 2$ we have:
    \\
\smallskip
    \\
$ \sum_{x \in X^{ \alpha}_{n}} \frac{T^{m}(x)}{2.5 \times m^{m+1}
\times f^{m}(x)} \times \mu(x) $ 
= $ \sum_{{\cal K} \subseteq 2^{n}} 
\sum_{x \in X^{ \alpha}_{n} \cap Th({\cal K})} \frac{T^{m}(x)}
{2.5 \times m^{m+1}\times f^{m}(x)} 
\times \mu(x) $ =
    \\
\medskip
    \\
= $ \sum_{{\cal K} \subseteq 2^{n}}  
\sum_{x \in X^{ \alpha}_{n} \cap Th({\cal K})} 
\frac{(min_{n}({\cal K}) + 1)^{m}}{2.5\times m^{m+1}} \times \mu(x) $ 
= $ \sum_{{\cal K} \subseteq 2^{n}}  
 \frac{(min_{n}({\cal K}) + 1)^{m}}{2.5\times m^{m+1}} 
\sum_{x \in X^{ \alpha}_{n} \cap Th({\cal K})} \mu(x) $ =
    \\
\medskip
   \\
= $ \sum_{{\cal K} \subseteq 2^{n}} \frac{(min({\cal K}) + 1)^{m}}{2.5\times m
^{m}}
\times \mu (X^{ \alpha}_{n} \cap Th({\cal K})) $ 
= $ \sum_{{\cal K} \subseteq 2^{n}}  
\frac{(min({\cal K}) + 1)^{m}}{2.5\times m^{m+1}}
 \times \frac{ \mu(X^{ \alpha}_{n})}
{2^{2^{n}}} $ = 
    \\
\medskip
   \\
= $ \frac{ \mu(X^{ \alpha}_{n})}{2.5\times m^{m+1}} \times \sum_{{\cal K} \subseteq 2^{n}} 
\frac{(min_{n}({\cal K}) + 1)}{2^{2^{n}}} $ 
= $ \frac{ \mu(X^{ \alpha}_{n})}{2.5\times m^{m+1}} \times \sum^{2^{n}}_{i=1}  
\frac{i^{m} \times 2^{2^{n}-i}}{2^{2^{n}}} $
$ \leq \frac{ \mu(X^{ \alpha}_{n})}{2.5\times m^{m+1}}
 \times \sum^{ \infty}_{i=1}  
\frac{i^{m}}{2^{i}}$.
   \\
\medskip
    \\
On the other hand,
$\sum^{ \infty}_{i=1}  
\frac{i^{m}}{2^{i}} = \sum^{m}_{i=1} \frac{i^{m}}{2^{i}} + 
 \sum^{\infty}_{i=m+1} \frac{i^{m}}{2^{i}}
\leq  \sum^{m}_{i=1} \frac{m^{m}}{2^{i}} +
 \sum^{\infty}_{i=m+1} (\frac{i}{2^{\frac{i}{m}}})^{m} =$ 
    \\
\medskip
   \\
$= m^{m}\times   \sum^{m}_{i=1} 2^{-i} +  
\sum^{\infty}_{\xi = \frac{m+1}{m},\Delta \xi = \frac{1}{m}} 
(\frac{\xi \times m}{2^{\xi }})^{m} \leq
m^{m} + m^{m} \times 
\sum^{\infty}_{\xi = \frac{m+1}{m},\Delta \xi = \frac{1}{m}} 
(\frac{\xi }{2^{\xi }})^{m} \leq $
    \\
\medskip
   \\
$ \leq m^{m} \times ( 1 + m \times \sum_{k=1}^{\infty} 
(\frac{k}{2^{k}})^m 
 \leq m^{m} \times ( 1 + m \times \sum_{k=1}^{\infty} 
\frac{k}{2^{k}} \leq \frac{1}{2} \times m^{m+1} + 2
\times m^{m+1} \leq $     \\
\medskip
   \\
$2.5 \times m^{m+1}$.

\medskip

Hence $ \sum_{x \in X^{ \alpha}_{n}} \frac{T^{m}(x)}{2.5 \times m^{m+1}
\times f^{m}(x)} \times \mu(x) \leq 
 \mu(X^{ \alpha}_{n})$, i.e.
 $ (T^{m})^{f, \mu}_{avg, \alpha} \in 
O(2.5 \times m^{m+1} \times \bullet ^{m}) $.

\bigskip    

There  is  a surprising (please take into account approximate calculations)
coincidence  between  the  constant  197 for 3-coloring backtrack search of
\linebreak
\cite{ Wilf84backtrack:an},  page 216, and the constant $3^{3} +
2 \times 3^{3+1} = 189$
of our estimation.

\section{A grain of salt}
\label{Grain}
\hspace*{\parindent}

\medskip    

As  we  have  seen  in  two  previous  sections,  under  rather  acceptable
assumptions  we  calculated  that  the  expected running time of tabulating
algorithm  does not exceed the cube of the time needed for merely rewriting
the  input, and that the expected running time of satisfiability testing is
less  than three times greater than the time spent on reading the input. Those result
may  or  may  not  hold  for  other  probability distributions. Despite its
seemingly naturalness,  the assumption of section \ref{Tabulating} we have made about
$  \mu(Y^{n}_{i})  $ is rather strong; as a matter of fact, it implies that
the  probability  of  a sentence decreases exponentially with the number of
distinct  variables  it  contains. (Here Shannon's counting argument fights
back).  In  our opinion it cannot be precluded that it is the most likely
probability  distribution  in  Artificial  Intelligence applications, where
verified  sentences  are  rather  far from being random in a lexical sense.
However,  if  we assume, that the probability $ \mu$ decreases with $p$-th
power  of  input's  length  then  the  following  example shows that $ T^{
\mu}_{avg}(X) = \infty $.

\begin{ex}
\rm
\label{Example 2}

 Consider a language containing all and only 16 binary connectives (i.e.
names of binary Boolean functions). Elementary calculations show that there
are 
\[ \Gamma (N) \times 16^{N} \times (2^{N+1} - 1) \]
 different sentences containing
exactly $N$ connectives (and therefore $ N + 1 $ propositional variables;
the set $V$ of this variables we treat as fixed here), where $ \Gamma(N) $
is defined inductively:     \\

$ \Gamma(0) = 1$,

\medskip

$ \Gamma(n + 1) = \sum^{n}_{i=0} \Gamma(i) \times \Gamma(n-i) $,
    \\
\bigskip
    \\
and  denotes  the  number of different types of sentences one may construct
out of $N$ binary connectives. Factor $ 2^{N+1} - 1 $ is the number of
possible selections from $V$.

\bigskip

Total time of reading all these sentences is equal to 
$ (2N + 1) \times \Gamma(N) \times 16^{N}(2^{N+1} - 1) $, while total
time of their tabulating is $ (2N + 1) \times \Gamma(N) \times 
16^{N} \times (3^{N+1} - 1) $. Therefore the ratio
$ F(N) = \frac{3^{N+1} -1}{2^{N+1} -1} \times (2N + 1)^{-p} \approx
(1.5)^{N+1} \times (2N + 1)^{-p} $ cannot have the convergent sum, i.e.
$ \sum^{ \infty}_{N=0} F(N) = \infty $. 

\bigskip

The  same  is  true  if  we assume, that input's probability decreases with
$p$-th power of the number of its propositional variables. \hfill $\Box$

\end{ex}

The situation becomes diametrically different if one assumes to have in the
language  all  possible  $n$-ary  connectives  for  each  $ n<0$, with fair
distribution  of  probability  over  arity  classes.  This  means that each
$n$-ary Boolean function has in this language its individual name which may
appear  in  input  equally  likely  with  any other name of $n$-ary Boolean
function.  The  explosion  of connectives and lengths of their codes should
substantially  contribute  to  the  enhancement of average relative running
time  of  tabulating  program: one may easily verify than assumption that $
\mu$ is constant on $ Y^{n}_{i} $ is satisfied in this case.

\bigskip

The  situation with the satisfiability problem is, hopefully, not as clear,
because  we did not use Shannon's counting argument here. Of course, having
all  possible  and equally likely connectives in a language forces that the
assumption   of   $  \mu(X^{  \alpha}_{n}  \cap  Th({\cal  K}))  =  \mu(X^{
\alpha}_{n}  \cap Th({\cal L})) $ is met. The more problematic case, where,
say, the arity of connectives is bounded, e.g. it cannot exceed 2, requires
further  investigation.  The answer to this problem is, probably, hidden in
the following question:

\smallskip

\begin{quote}

Assuming  that  all  and only $N$-ary connectives are present in the object
language,  and  that  any  two  sentences  of the same length have the same
probability, given number $M$, what is the expected value of $ min_{ \alpha
(x)}({\cal K}(x)) $, where $x$ is a random element of $ X^{f}_{M} $?

\end{quote}

\medskip    

\section{A comparison of methods}
\label{comparison}
\hspace*{\parindent}

\medskip    

In  our  opinion,  the  expected  complexity  $ T^{ \mu}_{avg} (X) $, and in
particular  its  finiteness,  is  the  most  adequate  complexity  measure,
provided  $P$  is intended for frequent future use, and the distribution of
probability $ \mu$ really describes what is going on in its input. The role
of  other  characteristics,  like  $  T^{f}_{x}, \: T^{f, \mu}_{avg},$ or $
T^{f,  \mu}_{avg,  \alpha} $, as well as asymptotic measures of complexity,
is  secondary,  as  they serve as a calculational facility in estimating
the  value  of  $  T^{  \mu}_{avg}(X)  $.  Incidentally,  the  knowledge of
worst-case  or  average  running  time  in the classic sense, or at least some
$O$-class  to  which  it  belongs,  may  be  sufficient to prove that $ T^{
\mu}_{avg}(X)  < \infty $, using e.g. Property \ref{Property 1}, but, as we
have  seen,  not  necessarily  in  all cases. On the other hand, a peculiar
conviction  that  O$(\bullet)$  is  much  better  than O$(2^{\bullet}) $ in
circumstances when the probability that in the next run the input will have
given length decreases with its second power, seems like preferring rain to
mud: both of them cause nontractability problems.

\bigskip

If  one  insists on having a characterization of how an increase in size of
input  space  would  affect  the  tractability  of  an  algorithm, Property
\ref{Property 1} is a neat tool for the purpose. It may be useful, e.g,. for
finding  a  maximal  $N$  such  that $ T^{ \mu}_{avg} ( \cup_{i \leq N} X^{
\alpha}_{i})  \leq  c  $,  where  $c$ is a limit of one's average patience.
Since,  on general, values of $ T^{ \mu}_{avg} ( \cup_{i \leq N} X^{f}_{i})
$   and  $  T^{  \mu}_{avg}  (X^{f}_{N})  $  may  differ  from  each  other
considerably,  using  to  this end the classical concept of average running
time,  besides  some  unnecessary  calculational problems which result from
restricting  $  \alpha$  to  $f$, may lead to false conclusions. Obviously,
asymptotic  measures  may  be impractical in such a case, since $N$ we
are interested in may be not large enough, i.e. less than $n_{0}$ appearing
in the definition of $O$-class.

\bigskip

Asymptotic  measures  may be adequate iff the probability of inputs of some
small  size  is  appropriately  small,  which would probably happen in most
cases where probabilities of any two inputs, or at least of any two input's
lengths,  were the same. However, if the input space is infinite, then such
distribution of probability is impossible, since in this case

\smallskip

\[ \mu(X) = \sum_{x \in X} \mu (x) = \left\{ \begin{array}{ll}
\sum_{x \in X} 0 = 0 \neq 1, & \mbox{if $ \mu(x)=0$}     \\
\sum_{x \in X} \varepsilon = \infty \neq 1, & \mbox{otherwise.}
\end{array}
\right. \] 

\smallskip

In our opinion the above fact is one of the reasons for discrepancies between
asymptotic and actual behaviors of many algorithms.

\bigskip

Using a  worst-case  measure in estimating algorithm efficiency is equivalent
to  average  case if the probability of non-worst inputs vanishes. This is
true under, as we call it, the malicious gnome assumption.

\medskip

\section{Final remarks}
\label{Conclusions}
\hspace*{\parindent}

\medskip    

Many  people  are  quite  skeptical  about  adequacy of probability theory,
seemingly  expecting  somebody  to  demonstrate  the ``truthfulness" of its
axioms.  We do not share their reservations, consciously leaving the choice
of  pertinent probability measure to lucky guessing of the applier. It does
not  mean,  however, that we see the results obtained on the ground of this
theory  as  nothing  but  speculations.  In  particular, we have found it a
little bit surprising, nevertheless instructive, that under quite realistic
assumptions a simple reading program may need, on average, as much as 30 \%
of  the running time of a satisfiability checker. This is why we wrote this
paper.

\bibliography{ref.bib}

\end{document}